\documentclass[10pt, aps, twocolumn, prd, showpacs, floatfix]{revtex4-2}

\usepackage{silence}
\usepackage{amssymb}
\usepackage{booktabs}
\usepackage{graphicx}
\usepackage{amsmath}
\usepackage{dcolumn}
\usepackage{bm}
\usepackage{soul}
\usepackage{comment}
\usepackage{xcolor}
\usepackage{hyperref}
\usepackage{cleveref}
\hypersetup{
    colorlinks=true,
    citecolor=blue,
    linkcolor=black,
    urlcolor=blue
    } 

\DeclareMathOperator{\HeunG}{HeunG}

\begin{document}

\preprint{APS/123-QED}

\title{Exact Solution of the Non-minimally Coupled Klein-Gordon Equation in the Schwarzschild Star}
\author{Reynan A. Dulinayan}
\email{radulinayan@nip.upd.edu.ph}
\affiliation{National Institute of Physics, University of the Philippines Diliman}

\author{Kevin T. Grosvenor}
\email{kgrosvenor@nip.upd.edu.ph}
\affiliation{National Institute of Physics, University of the Philippines Diliman}

\vspace{-15pt} 
\date{\today}

\date{\today}

\begin{abstract}
We present for the first time the exact solution of the massive Klein-Gordon equation in the Schwarzschild star (perfect-fluid, uniform-density, spherically-symmetric star), including the non-minimal curvature-scalar coupling. The solution is expressed in terms of the general Heun function. A geometry-induced algebraic coordinate transformation reveals a hidden Fuchsian structure that underlies the exact solvability. Known leading- and next-to-leading-order results are recovered in the low-compactness limit. In the Buchdahl limit, we derive a regularity condition for static modes and describe analytically the divergence in amplitude and oscillation wave vector of dynamic modes as they approach the pressure singularity at the center of the star.
\end{abstract}

\maketitle

\tableofcontents

\section{Introduction}

The Schwarzschild star metric is an analytic solution in General Relativity of great foundational importance in modeling compact objects \cite{Schwarzschild_1916_Interior, Lattimer:2000nx, Koliogiannis:2024szd}. Scalar field perturbations on this background have been studied intensely from the perspective of stellar superradiance \cite{Cardoso:2015zqa}, scattering \cite{Stratton2017, OuldElHadj:2019kji}, magnetars \cite{Dariescu_2017} and quantum field theory \cite{gossel_2011, Reyes:2023fde, Flambaum:2012yw}. The central challenge is to solve the Klein-Gordon equation in the stellar interior and to match this at the stellar surface with the exterior solution. This is important for dark matter capture in the interior of stars \cite{Deliyergiyev:2023uer, Brito:2015yga, Guver:2012ba, Bramante:2015cua}, for instance via multiscattering events \cite{Bramante:2017xlb}.

One of the leading candidates for dark matter is an ultralight boson \cite{Turner:1983he, Press:1989id, Sin:1992bg, vandeBruck:2022xbk} described by a minimally-coupled massive Klein Gordon field \cite{Hui:2016ltb, Hu:2000ke, Sikivie:2009qn, Foidl:2022bpn, Kousha:2023jzt}. Despite being a primary theoretical baseline, as of yet, there is no known exact analytic solution to the Klein Gordon equation in the Scwharzschild star metric. Instead, the aforementioned works rely on numerical techniques and approximations. In contrast, analytic solutions for field equations in standard black hole backgrounds are well-established and frequently given in terms of general or confluent Heun functions \cite{Suzuki:1998vy, Fiziev2006, Fiziev2010, Bezerra2014}. These solutions have become benchmarks for subsequent applications to quasi-normal modes \cite{Fiziev:2011mm}, black-hole resonant frequencies and Hawking radiation \cite{Vieira:2014waa, Vieira:2016ubt}, spacetime interpretations \cite{Minucci:2024qrn} and superradiant instabilities \cite{Baybay:2025kvb}.

In this paper, we provide for the first time the exact solution to the massive Klein Gordon equation in the Schwarzschild star background, expressed in terms of the general Heun function. The solution captures the Minkowski limit and the small-compactness approximation made by Ref.~\cite{Dariescu_2017}. Crucially, the exact solution allows us to map out precisely the behavior of the scalar field in the strong gravity limit, isolating the effects of extreme background curvature.

At the core of this exact solvability is an algebraic coordinate transformation introduced by Schwarzschild \cite{Schwarzschild_1916_Interior} which maps the stellar interior to a spatial 3-sphere geometry. This is the same transformation that Petroff \cite{Petroff:2007tz} used to put the frame-dragging equation for a slowly rotating constant-density star into the general Heun form. It is also the transformation used by Horbatsch and Burgess \cite{Horbatsch:2010hj} to solve a particular limit of the Brans-Dicke scalar field equation in this background geometry in terms of the general Heun function. Unlike the usual M\"obius transformations used only to canonically fix the locations of three of the singularities of the radial equation (at $0$, $1$, and $\infty$), the algebraic transformation linearizes the metric coefficients, reorganizes the singularity structure, and exposes a hidden Fuchsianity, which ultimately leads to the exact solution. 


\section{Klein-Gordon field in the Schwarzschild star}

We consider the action
\begin{align}
    S = - \frac{1}{2} \int d^4 x \sqrt{-g} \Bigl( \nabla^{\nu} \Phi^* \nabla_{\nu} \Phi + (\mu^2 + \xi R) |\Phi|^2 \Bigr),
\end{align}
and the resulting field equation
\begin{align}
    ( \Box - \mu^2 - \xi R ) \Phi = 0,
\end{align}
where $\xi$ is the non-minimal coupling constant. We work in geometric units $G = \hbar = c = 1$. The stellar mass and radius are $M$ and $R_{\star}$, respectively. We define dimensionless temporal and radial coordinates, field mass, spacetime curvature, and effective compactness $\kappa$:
\begin{align}
\{ \bar{t}, \bar{r} \} &= \frac{\{ t, r \}}{R_{\star}}, &%
\{ \bar{\mu}^2 , \bar{R} \} &= \{ \mu^2 , R \} R_{\star}^{2}, &%
\kappa &= \frac{2M}{R_{\star}},
\end{align}
where the extra factor of 2 compared to the standard definition of the compactness parameter is introduced solely for convenience.

The Schwarzschild star is sourced by a constant-density, incompressible perfect fluid, and the line element $ds^2$ is given by
\begin{subequations} \label{eq:TheMetric}
\begin{align}
    \tfrac{1}{R_{\star}^2} ds^2 &= - N ( \bar{r} )^2 d \bar{t}^2 + \tfrac{1}{f( \bar{r} )} d \bar{r}^2 + \bar{r}^2 d \Omega^2, \label{eq:metric} \\
    N( \bar{r} ) &= \tfrac{3}{2} \sqrt{1 - \kappa} - \tfrac{1}{2} \sqrt{1 - \kappa \bar{r}^2}, \label{eq:theN} \\
    f( \bar{r} ) &= 1 - \kappa \bar{r}^2. \label{eq:thef}
\end{align}
\end{subequations}
As is standard, we assume the probe limit and ignore the backreaction of the Klein-Gordon field on the background spacetime. The field equation on this background is separable and the solution can be written as
\begin{equation} \label{eq:sep}
\Phi ( \bar{t} , \bar{r} , \vartheta, \varphi) = \Psi ( \bar{r} ) e^{-i \bar{\omega} \bar{t}} Y_{\ell}^m (\vartheta, \varphi).
\end{equation}
The radial equation reads
\begin{equation} \label{eq:KGradial}
\frac{d^2 \Psi}{d \bar{r}^2} + P \frac{d \Psi}{d \bar{r}} + Q \Psi = 0,
\end{equation}
where
\begin{subequations}
\begin{align}
P &= \frac{d}{d \bar{r}} \ln \bigl( \bar{r}^2 N \sqrt{f} \bigr), \\
Q &= \frac{1}{f} \biggl( \frac{\bar{\omega}^2}{N^2} - \bar{\mu}^2 - \xi \bar{R} -  \frac{\ell ( \ell + 1 )}{\bar{r}^2} \biggr). \\
\bar{R} &= 3 \kappa ( 1 - 3 w ),
\end{align}
\end{subequations}
where $w$ is the equation of state, given by the ratio of pressure and density,
\begin{equation}
    w = \frac{p}{\rho} = \frac{\sqrt{1 - \kappa \bar{r}^2} - \sqrt{1 - \kappa}}{3\sqrt{1-\kappa} - \sqrt{1- \kappa \bar{r}^2}}.
\end{equation}
Previous work has sought to solve Eq.~\eqref{eq:KGradial} either numerically or approximately. For example, in the minimally coupled regime ($\xi = 0$), Cardoso et al. \cite{Cardoso:2015zqa} and Dariescu et al. \cite{Dariescu_2017} both work in the small compactness limit. The former sets the compactness to zero, in which case the solutions are Bessel functions, while the latter demonstrates that, to first order in compactness, the solution can be expressed in terms of the general Heun function. We claim that this equation is solved \emph{exactly} using the general Heun function. The solution expanded to zeroth-order in $\kappa$ reproduces the flat-space result, and expanded to first-order in $\kappa$ is equivalent to the result in \cite{Dariescu_2017}.

This finding is significant for several reasons. Firstly, we no longer need to rely on numerics to solve this important baseline problem. Any discrepancies between numerics and analytics can be addressed more reliably (see \cite{Baybay:2025kvb} for an example of this in the case of the superradiant instability of Kerr-Newman black holes). Secondly, the analysis of perturbations of this setup can be validated more reliably since the backgrounds are exact. 


\section{General Heun mapping}

Using the aforementioned algebraic coordinate transformation as in Ref.~\cite{Petroff:2007tz, Horbatsch:2010hj},
\begin{equation}\label{eq:trans}
x = \tfrac{1}{2} \bigl( 1 - \sqrt{1 - \kappa \bar{r}^2} \bigr),
\end{equation}
the radial equation \eqref{eq:KGradial} becomes
\begin{equation} \label{eq:Psieq}
\Psi'' + \hat{P} \Psi' + \hat{Q} \Psi = 0,
\end{equation}
where $'$ denotes differentiation with respect to $x$, and $\hat{P}$ and $\hat{Q}$ are functions of $x$ with poles at $x=0$, $x=1$, and $x=a$, with the latter given by
\begin{equation} \label{eq:thea}
a \equiv \tfrac{1}{2} - \tfrac{3}{2} \sqrt{1 - \kappa}.
\end{equation}
The poles at $x=0$ and $x=1$ are both at the stellar center ($\bar{r} = 0$), but on different branches of the square root function. The pole at $x=a$ corresponds to the values $\bar{r} = \bar{r}_{\pm} \equiv \pm 3 \sqrt{1 - \frac{\kappa_B}{\kappa}}$, where $\kappa_B = \frac{8}{9}$ is the Buchdahl limit \cite{Buchdahl_1959}. The equation of state parameter is $w = \frac{1+a - 3x}{3(x-a)}$, and thus, the pressure diverges at $x=a$. However, since $-1 \leq a < 0$, this location is unphysical. The upper limit, $a=0$, corresponds to a pressure singularity at the stellar center at the Buchdahl limit \footnote{Schwarzschild \cite{Schwarzschild_1916_Interior} showed that $\frac{M}{R} < \frac{4}{9}$ and that the internal pressure diverges at the origin when $\frac{M}{R} = \frac{4}{9}$. This was generalized by Buchdahl \cite{Buchdahl_1959} to a ball of any stable, static and barotropic perfect fluid. For this reason, the limit bears Buchdahl's name.}. Since $0 \leq \bar{r} \leq 1$ and $0 \leq \kappa < \frac{8}{9}$, the physical range of $x$ is $0 \leq x \leq \frac{1}{3}$, and so the pole at $x=1$ is also at an unphysical location. Nevertheless, these singularities are essential to the form of Eq.~\eqref{eq:Psieq} as the functions $\hat{P}$ and $\hat{Q}$ are given by
\begin{subequations} \label{eq:PQhat}
\begin{align}
    \hat{P} & = \tfrac{3}{2x} - \tfrac{3}{2(1-x)} + \tfrac{1}{x-a} \label{eq:friction1} \\
    \hat{Q} & =  \tfrac{E_0}{x} +  \tfrac{E_1}{1-x} + \tfrac{E_a}{x-a} + \tfrac{F_0}{x^2} + \tfrac{F_1}{(1-x)^2}   + \tfrac{F_a}{(x-a)^2}, \label{eq:potent1}
\end{align}
\end{subequations}
where, for $n=0$ and $1$,
\begin{subequations} \label{eq:EF}
\begin{align}
E_n &= \tfrac{\bar{\omega}^2}{\kappa (n-a)^2} - \tfrac{\bar{\mu}^2}{\kappa} - \tfrac{\ell(\ell+1)}{2} - 12 \xi - \tfrac{ 3\xi(2a-1)}{n-a} \\
E_a &= E_1 - E_0 = \frac{\bar{\omega}^2 (2a - 1)}{\kappa a^2 (1-a)^2} - \frac{3 \xi (2a-1)}{a (1-a)}, \\
F_0 & = F_1 = - \frac{\ell(\ell+1)}{4}, \\
F_a &= \frac{\bar{\omega}^2}{\kappa a(1-a)}.
\end{align}
\end{subequations}
The solution behaves as a power law near each pole with exponents given by solutions to the indicial equations. We peel these asymptotics off, writing the solution as
\begin{equation} \label{eq:shomotopy}
    \Psi (x) = x^{\sigma_0} (1-x)^{\sigma_1} (x-a)^{\sigma_a} \mathcal{R} (x),
\end{equation}
with
\begin{align} \label{eq:sigmaa}
    \sigma_0 & =  \sigma_1  = \frac{\ell}{2}, & 
    \sigma_a & = - \frac{\bar{\omega}}{\sqrt{\kappa a (a-1)}}.
\end{align}   
The equation for the function $\mathcal{R}$ reads
\begin{align} \label{eq:ode3}
    \mathcal{R}'' + \bigl( \tfrac{\gamma}{x} + \tfrac{\delta}{x-1} + \tfrac{\varepsilon}{x-a} \bigr) \mathcal{R}'  + \tfrac{\alpha \beta x - q}{x (x-1)(x-a)} \mathcal{R} = 0  ,
\end{align}
where
\begin{subequations}
\begin{align}
    \gamma & =
    \delta  = \ell + \tfrac{3}{2}, \label{eq:delta} \\
    \varepsilon & = 2 \sigma_a + 1, \label{eq:varepsilon} \\
    \genfrac{\{}{\}}{0pt}{}{\alpha}{\beta} & = \sigma_a + \ell + \frac{3}{2} \pm \sqrt{\frac{9}{4} - \frac{\bar{\mu}^2}{\kappa} - 12 \xi}, \label{eq:alphas}  \\
    q & = - \sigma_a^2 (a-1) + \sigma_a \bigl( \ell +\tfrac{3}{2} \bigr)+ a(\ell+1)^2 + \tfrac{\ell}{2} \notag \\
    &\quad - a \bigl( 1 - \tfrac{\bar{\mu}^2}{\kappa} \bigr) + 3 \xi \bigl( 2a + 1 \bigr). \label{eq:accesory}
\end{align}
\end{subequations}
This is the general Heun equation \cite{Heun_1889, Ronveaux_1995}, provided that the following Fuchsian relation hold
\begin{align} \label{eq:Fuchsianidentity}
\gamma + \delta + \epsilon = \alpha + \beta + 1,
\end{align}
which is easily verified. There are, of course, two linearly independent general Heun functions, but only one is regular at $x=0$, namely
\begin{equation} \label{eq:calR}
\mathcal{R} (x) = \mathcal{R}_0 \HeunG ( a, q; \alpha , \beta, \gamma, \delta; x), 
\end{equation}
where $\mathcal{R}_0 = \mathcal{R} (0)$ and where we have used the standard normalization $\HeunG ( a , q ; \alpha , \beta , \gamma, \delta ; 0) = 1$.

Note that $\bar{\mu}$ and $\xi$ drop out of the Fuchsian relation. This demonstrates that the exact solvability of the wave equation on the Schwarzschild star background is intrinsic to the background itself, not to the field. 

In fact, the solvability of the scalar field in this background in terms of the general Heun function is not limited to standard non-minimal coupling. Additional non-minimal couplings may be added up to second-order in curvature in the potential term and up to first-order in curvature in the kinetic term as long as the couplings satisfy various constraints that enforce that no new singularities are generated.


\section{The 3-sphere geometry}

Eq. \eqref{eq:trans} can be interpreted as a half-angle transformation,
\begin{align} \label{eq:HalfAngle}
x = \sin^2 ( \chi / 2 ),
\end{align}
where $\cos \chi = \sqrt{1 - \kappa \bar{r}^2}$, which is well defined since $ 0 \leq \kappa \bar{r}^2 < \frac{8}{9}$. In angular coordinates $( \chi , \vartheta, \varphi )$, the spatial part of the line element \eqref{eq:metric}, denoted by $dl^2$, reads
\begin{align}
    dl^2 = \tfrac{1}{\kappa} \left[ d \chi^2 + \sin^2 \chi \left( d\vartheta^2 + \sin^2 \vartheta \, d \varphi^2 \right) \right],
\end{align}
which is the metric of a 3-sphere of radius $1/ \sqrt{\kappa}$. The range of the hyperpolar angle is $0 \leq \chi \leq \cos^{-1}(\sqrt{1-\kappa})$, and so the space is, in fact, a hyperspherical cap. As previously mentioned, this structure was first pointed out by Schwarzschild \cite{Schwarzschild_1916_Interior} in his original derivation. Recent work \cite{Seenivasan:2025ysy} leverages this connection to obtain closed-form expressions for mode solutions of the conformally coupled scalar ($\xi = \frac{1}{6}$ and $\mu = 0$). The 3-sphere structure suggests expanding the solution in terms of hyperspherical harmonics. Nevertheless, the nontrivial dependence of the lapse function \eqref{eq:theN} on $\bar{r}$, and thus on $\chi$, ultimately leads to a $\chi$-equation of general Heun type. This provides an alternative, but equivalent, argument for the exact solvability.


\section{Weak gravity regime}

The Schwarzschild interior becomes flat Minkowski spacetime when the compactness is set to zero. The non-minimal coupling term also vanishes in this limit. In this case, the solution of the radial equation is a spherical Bessel function. Indeed, in this limit, the three-term recurrence relation of the general Heun function \cite{Maier2007} reduces to a two-term one (see App.~\hyperref[app:A]{A}), which is easily solved to yield the Taylor expansion around $\bar{r}=0$, 
\begin{align} \label{eq: Rsmallkappa}
    \mathcal{R}(x) \xrightarrow{\kappa \rightarrow 0} \sum_{k=0}^{\infty} \frac{(-1)^k(2\ell + 1)!!}{2^k k!(2\ell + 2k +1)!!} [(\bar{\omega}^2 - \bar{\mu}^2) \bar{r}^2]^k. 
\end{align}
We recognize the spherical Bessel function in this expansion. The full radial function becomes
\begin{align}
    \Psi(r) \xrightarrow{\kappa \rightarrow 0} \mathcal{C} \, \frac{ \kappa^{\ell/2} (2\ell+1)!!}{2^{\ell} \left(\bar{\omega}^2 - \bar{\mu}^2 \right)^{\ell/2}} \, j_{\ell} \left( \sqrt{\omega^2 - \mu^2} \, r \right),
\end{align}
where $\mathcal{C}$ is a constant coefficient.

\begin{figure}[t!]
    \centering
    \includegraphics[width=1\linewidth]{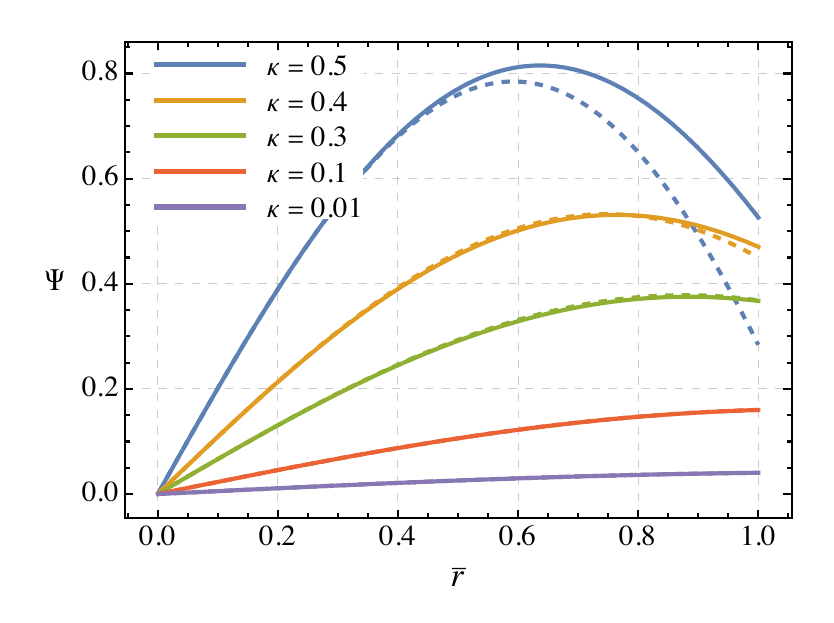}
    \caption{Comparison of the exact solution, \eqref{eq:shomotopy} and \eqref{eq:calR} with \eqref{eq:trans} (solid curves) and the approximation solution \eqref{DariescuHeun} of \cite{Dariescu_2017} at different values of compactness $\kappa$. The parameters are $\xi = 0$, $\bar{\mu} = 1$, $\bar{\omega} = 2$, and $\ell = 1$.}
    \label{VersusDariescu}
\end{figure}

Expanding the background to linear order in $\kappa$, Ref.~\cite{Dariescu_2017} obtained an approximate solution to Eq. \eqref{eq:KGradial},
\begin{align} \label{DariescuHeun}
    \Psi( \bar{r} ) \approx  \mathcal{C} \bar{r}^l \text{HeunG} (a',q'; \alpha',\beta',\gamma',\delta'; \zeta), 
\end{align}
where $\mathcal{C}$ is a constant, $\zeta = 3 \kappa \bar{r}^2/(3\kappa+4)$, and the primed parameters are given in their Eq.~22. Strictly speaking, this solution is not quite correct since it contains higher orders in $\kappa$ even though such higher-order terms were ignored in the background. Nevertheless, it is correct to linear order in $\kappa$. Furthermore, an analysis of the recurrence relation in the strict $\kappa \rightarrow 0$ limit reveals the same spherical Bessel function result as our exact solution (see App.~\hyperref[app:A]{A}). Finally, we show the consistency between this approximate solution (dashed) and our exact solution (solid) in Fig.~\ref{VersusDariescu} (turning off their magnetic field and our $\xi$ parameter while setting $\mathcal{C} = 1$ for both solutions to facilitate the comparison). The agreement is excellent at low to moderate compactness values with deviations emerging at high compactness, as anticipated. Since these deviations rapidly grow towards the stellar surface, using the approximate solution to address the delicate problem of matching the interior and exterior solutions at $\bar{r} = 1$ is impracticable and is another reason why the exact solution is so valuable.


\section{Strong gravity regime}

We now investigate the behavior of the scalar field when the compactness is near the Buchdahl limit $\kappa_B = 8/9$. A star with this compactness is a possible stable configuration after stellar gravitational collapse \cite{Shaymatov:2022hvs, Shaymatov:2022ako,Alho:2022bki, Chakrabarti:2026var} and is a premiere candidate \textit{black hole mimicker}, a horizonless ultra-dense compact object that behaves like a black hole \cite{Dadhich:2022yuk}. In fact, ultra-compact objects with compactness in the regime $2/3 \leq \kappa < 8/9$ are already capable of developing a light ring \cite{Urbano_2019, Raposo:2018rjn, Cardoso2017Nature, Cardoso2019LRR} and are considered mimicker candidates. Neglecting backreaction in this regime is clearly a limitation \cite{Reyes:2023fde}, as it may affect the Buchdahl bound \cite{Arrechea:2023oax}. Nevertheless, it is valuable to isolate the effects of extreme background curvature on the behavior of the scalar field.


\subsection{Static modes}

Setting $\bar{\omega} = 0$ and $\kappa \rightarrow \kappa_B$ or, equivalently, $a \rightarrow 0$, turns Eq.~\eqref{eq:ode3} into an equation with a second-order pole at $x=0$. We can remove this second-order pole by writing $\mathcal{R} (x) = x^{\lambda} g(x)$, with 
\begin{equation}
    \lambda = - \frac{1}{2} \left( \ell + \frac{3}{2} \right) + \frac{1}{2} \sqrt{\left( \ell + \frac{3}{2} \right)^2 -4q},
\end{equation}
where $q = \frac{\ell}{2} + 3 \xi$ in this limit. The equation for $g$ may be written in the form
\begin{equation} \label{eq:2F1eq}
x(1-x) g'' + \bigl[ C - (1 + A + B)x \bigr] g' - ABg = 0,
\end{equation}
where the parameters $A$, $B$, and $C$ are 
\begin{align}
    A & = \lambda + \alpha, &
    B & = \lambda + \beta, &
    C & = 2 \lambda + \ell + \tfrac{5}{2},
\end{align}    
where $\alpha$ and $\beta$ are expressed as in Eq.~\eqref{eq:alphas}, but with $\sigma_a = 0$ and $\kappa = \frac{8}{9}$. The solution to Eq.~\eqref{eq:2F1eq} is of hypergeometric type where the regular part at $x=0$ is ${}_2 F_{1} (A,B; C; x)$. Thus, the functional form of the static mode near the Buchdahl limit has radial part
\begin{align} \label{eq:hypersol}
    \Psi(x) |_{\omega = 0} \xrightarrow{\kappa \rightarrow \kappa_B} \mathcal{C} \, x^{\frac{\ell}{2} + \lambda} (1-x)^{\frac{\ell}{2}}  {}_2F_1 \bigl( A , B ; C ; x \bigr),
\end{align}
for some constant $\mathcal{C}$ within the domain $0 \leq x \leq \frac{1}{3}$. Utilizing the half-angle transform,  Eq.~\eqref{eq:2F1eq} can be converted into a Schr\"odinger equation with a \emph{P\"oschl-Teller} potential (see App.~\hyperref[app:B]{B}). It is also well known that the solutions to this potential take the hypergeometric form \cite{PoschlTeller1933,daSilva:2023asj}.

The Schwarzschild star on the brink of developing a pressure singularity at the center can host regular static solutions at that point, provided that $\frac{\ell}{2} + \lambda \geq 0$ or, equivalently, $\ell ( \ell +1 ) \geq 12 \xi$. At minimal coupling, static modes with any $\ell$ remain regular. The $\ell = 0$ mode loses regularity for any $\xi > 0$, the $\ell = 1$ mode loses regularity above conformal coupling $\xi > 1/6$, etc. Intuitively, this phenomenon arises from a competition between the repulsive centrifugal potential and the attractive potential due to the coupling to the curvature, which is diverging at the center of the star.


\subsection{Dynamic modes}

\begin{figure}[t!]
    \centering
    \includegraphics[width=1\linewidth]{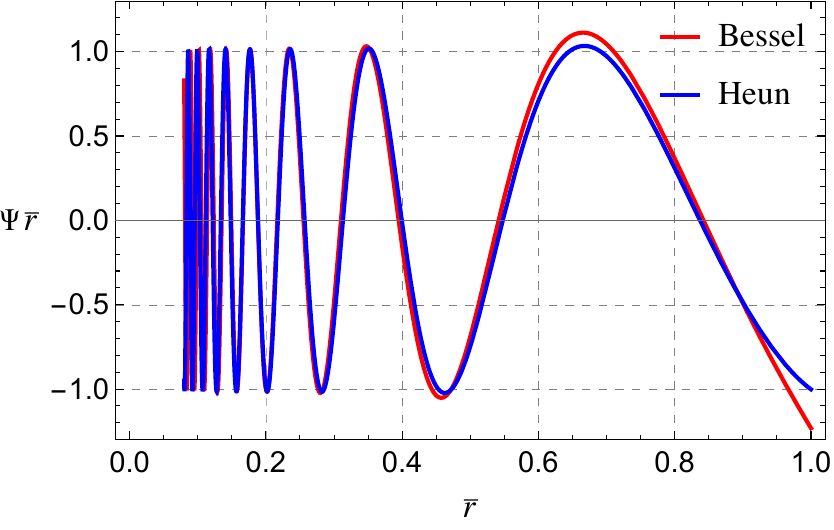}
    \caption{Plot of $\bar{r} \Psi (x)$ near the Buchdahl limit using the Bessel result Eq.~\eqref{eq:DynamicBuchdahl} (red) and the exact Heun solution evaluated at $\kappa = 0.8888$ (blue). The parameters are $\xi = 0$, $\bar{\mu} = 0$, $\bar{\omega} = 1$, and $\ell = 1$.}
    \label{fig:chirping}
\end{figure}

The scalar field diverges in the stellar interior near the Buchdahl limit. The field also exhibits a phenomenon we call \emph{chirping}: oscillating infinitely many times towards the center of the star. To see this, write $\mathcal{R} (x) = \exp \left(-\frac{1}{2} \int^{x} P(x') dx' \right) h(x)$ where $P$ is the friction term of Eq. \eqref{eq:ode3}. Extract the Schr\"odinger equation for $h(x)$. To capture the leading behavior as $a \rightarrow 0$ and near $x=0$, keep only the poles at $x=0$ and $x=a$, discarding those at $x=1$, and keep only the leading term in the coefficient at each pole. The resulting equation for $h(x)$ is
\begin{align} \label{eq:dHyper}
    h'' + \left( \frac{\bar{\omega}^2}{\kappa_B x(x-a)^2} + \frac{\frac{\gamma}{2} ( 1 - \frac{\gamma}{2} )}{x^2} \right) h = 0.
\end{align}
The solution for finite $a$ is ${}_2 F_{1}$, and $a \to 0$ is a \emph{confluent} limit, resulting in ordinary or spherical Bessel functions (see App.~\hyperref[app:C]{C}). To see this directly, change variables to 
\begin{equation}
z = 2 \bar{\omega} / \sqrt{\kappa_B x} 
\end{equation}
and set $y(z) = h(x) / x^{1/4}$. In the limit $a \rightarrow 0$, the equation for $y(z)$ becomes
\begin{align} \label{eq:BesselDE}
    z^2 \frac{d^2 y}{ dz^2} + 2z\frac{dy}{dz} + \bigl[ z^2 - \ell ( \ell + 1 ) \bigr] y = 0,
\end{align}
which is the spherical Bessel equation. The solution for $\Psi (x)$ with     dynamic modes near the Buchdahl limit is 
\begin{align}\label{eq:DynamicBuchdahl}
    \hspace{-0.1cm} \Psi(x) |_{\omega \neq 0}  \xrightarrow{\kappa \rightarrow \kappa_B} \frac{\mathcal{C}_1 j_{\ell} (z) + \mathcal{C}_2 y_{\ell} (z)}{x (1-x)^{3/4}},  
\end{align}
for some constants $\mathcal{C}_1$ and $\mathcal{C}_2$, where $j_{\ell}$ and $y_{\ell}$ are spherical Bessel functions of the first and second kind. By comparing with the ${}_2 F_1$ solution at finite $a$, we find that $\mathcal{C}_1$ and $\mathcal{C}_2$ diverge as $a \rightarrow 0$ (see App.~\hyperref[app:C]{C})\footnote{Note, however, that no physical significance can be ascribed to the divergence of the amplitude until a normalization condition is imposed.}.  Meanwhile, as we approach the stellar center, $z \rightarrow \infty$, and the spherical Bessel functions become sinusoidal with $2 \pi$ periodicity in $z$. As a function of $\bar{r}$, the field oscillates with increasing wave vector, or decreasing wavelength, towards the stellar center (chirping). In Fig.~\ref{fig:chirping}, we plot $\bar{r} \Psi$ using the exact Heun solution near the Buchdahl limit with $\kappa = 0.8888$ (blue) and using Eq.~\eqref{eq:DynamicBuchdahl} (red). We manually removed the divergence in the amplitude, effectively normalizing it, and tuned the constants $\mathcal{C}_1$ and $\mathcal{C}_2$ for the best match. This provides the first exact analytic derivation of the chirping phenomenon, which has been observed for $\ell  = 0$ modes of a minimally coupled massless scalar field using numerics and approximations \cite{Flambaum:2012yw}. 


Note that $T_{\bar{r} \bar{r}}^{\rm (fluid)} = p ( \bar{r} )$ diverges at the Buchdahl limit near $\bar{r} = 0$ like $\bar{r}^{-2}$. However, the most divergent contribution to $T_{\bar{r} \bar{r}}^{\rm (scalar)}$ is $( \frac{d \Psi}{d \bar{r}} )^2$ when the derivative acts on the spherical Bessel function, and behaves like $\bar{r}^{-6}$. Therefore, $T_{\bar{r}\bar{r}}^{\rm (scalar)}$ dominates over $T_{\bar{r}\bar{r}}^{\rm (fluid)}$ below a crossover radius where backreaction is no longer negligible. 
 

\section{Conclusions}

In this work, we showed that the massive Klein-Gordon equation non-minimally coupled to the Schwarzschild star background can be solved exactly in terms of the general Heun function. We capture the known results at minimal coupling in the zeroth- or first-order low-compactness limits. We also analyzed the behavior of the scalar field near the Buchdahl limit. We obtained exact static-mode solutions in terms of the hypergeometric ${}_2 F_1$ function, and their regularity condition in terms of the angular momentum number $\ell$ and non-minimal coupling $\xi$. We obtained dynamic-mode solutions in terms of spherical Bessel functions and derived the chirping phenomenon characterized by a diverging field amplitude as we approach the Buchdahl limit and a diverging oscillation wave vector as we approach the pressure singularity at the stellar center. Thus, the $rr$-component of the scalar-field stress tensor eventually exceeds the fluid pressure, signaling the breakdown of the probe limit.

This work completes the exact analytic picture of scalar field perturbations in the \emph{interior} Schwarzschild star background. The \emph{exterior} admits well-known solutions in terms of confluent Heun functions \cite{Fiziev2006, Vieira:2014waa}. The next step is to perform the matching at the stellar surface. This is a delicate procedure and is difficult to execute accurately using approximations or numerics. With our exact solutions, we are in a much better position. Nevertheless, this is still a big challenge and will likely require more tools (e.g., the conformal blocks approach \cite{Aminov:2020yma, Bonelli:2021uvf}). We hope to address this in future work.

Additionally, there are two clear directions for future work: more realistic stellar equations of state and backreaction. Having solved the important baseline problem of the Klein-Gordon field in a constant-density star completely analytically, we now have a solid base on which to study perturbations of this model.


\section*{Acknowledgments}

The authors acknowledge the Office of the Chancellor of the University of the Philippines Diliman, through the Office of the Vice Chancellor for Research and Development, for funding support through the Outright Research Grant (262608 ORG). 

\bibliography{references}

\clearpage

\appendix


\section{The Minkowski limit} 
\label{app:A}

The general Heun solution recovers the spherical Bessel solution in the flat space limit. The leading-order behaviors of the general Heun variables as $\kappa \rightarrow 0$ are
\begin{subequations} 
\begin{gather}\label{eq:minklimvars}
    x = \frac{1}{4} \kappa \bar{r}^2, \qquad a = -1, \qquad q = \frac{\bar{\omega}^2 - \bar{\mu}^2}{\kappa}, \\
    \genfrac{\{}{\}}{0pt}{}{\alpha}{\beta} = \frac{1}{\sqrt{\kappa}} \left( - \frac{\bar{\omega}}{\sqrt{2}} \pm i \bar{\mu} \right), \\
    \gamma = \delta = \ell + \frac{3}{2}, \qquad \varepsilon = -\frac{\sqrt{2} \bar{\omega}}{\sqrt{\kappa}} .
\end{gather}
\end{subequations}
The recursion relation between the coefficients $c_k$ of the Taylor expansion $\sum_{k = 0}^{\infty} c_k x^k$ of the general Heun function around $x = 0$ reads
\begin{subequations}
\begin{align}
    & A_k c_k + A_{k-1} c_{k-1} + A_{k-2} c_{k-2}  = 0, \\ 
    & A_k  = 2k ( k -1 + \gamma )a, \\
    & A_{k-1}  = -2(k-1) (k -2 + \gamma + \delta )a \\ 
    & \hspace{34pt} -\! 2(k-1) (k -2 + \gamma + \varepsilon )  -2 q, \\
    & A_{k-2} = 2 (k -2 + \alpha ) (k -2 + \beta ),
\end{align}
\end{subequations}
with $c_{-1} = 0$ and $c_0 = 1$. To leading order in $\kappa$,
\begin{subequations}
\begin{align}
    A_k &= - k (2\ell + 2k + 1), \\
    A_{k-1} &= - 2q = \frac{2(\bar{\mu}^2 - \bar{\omega}^2)}{\kappa}, \\
    A_{k-2} & = 2 \alpha \beta = \frac{2\bar{\mu}^2 + \bar{\omega}^2}{\kappa}.
\end{align}    
\end{subequations}
Since $A_{k-1}$ and $A_{k-2}$ are of the same order in $\kappa$, we can drop $A_{k-2}$ altogether from the recursion relation. This is proven by induction. The base case is given by fiat since $c_{-1} = 0$. Suppose that $c_k = O(\kappa^{-k})$ for $k \leq n$, where $n \geq 3$ is an integer. Consider the recursion relation at level $n+1$: $A_n = O(\kappa^{-1})$ and $c_n = O(\kappa^{-n})$, whereas $A_{n-1} = O(\kappa^{-1})$ and $c_{n-1} = O(\kappa^{1-n})$. Thus, $A_n c_n = O(\kappa^{-n-1})$, whereas $A_{n-1} c_{n-1} = O(\kappa^{-n})$ is subleading. This proves that the three-term recursion relation reduces to a two-term one in the $\kappa \to 0$ limit:
\begin{align}
    c_k = \frac{2}{k(2\ell + 2k +1)} \left( \frac{\bar{\mu}^2 - \bar{\omega}^2}{\kappa} \right) c_{k-1}.
\end{align}
Given $c_0 = 1$, the general solution is
\begin{align}
    c_k = \frac{2^k}{k!} \frac{(2\ell +1)!!}{(2\ell + 2k +1)!!} \left( \frac{\bar{\mu}^2 - \bar{\omega}^2}{\kappa} \right)^k.
\end{align}
Expressing $x$ in terms of $\bar{r}$ in this limit via Eq.~\eqref{eq:minklimvars}, derives the Taylor expansion Eq.~\eqref{eq: Rsmallkappa} of the general Heun function around $\bar{r} = 0$. This is related to the spherical Bessel function via the Taylor expansion of the latter,
\begin{align}
    j_{\ell} (x) = \frac{x^{\ell}}{(2\ell +1)!!} \sum_{k=0}^{\infty} \frac{(-1)^k}{2^k k!} \frac{(2\ell +1)!!}{(2\ell + 2k + 1)!!} x^{2k}.
\end{align}
The same argument shows that the small compactness approximation of Dariescu et al. \cite{Dariescu_2017} gives the same spherical Bessel function in the limit $\kappa \to 0$. In this limit, their parameters read
\begin{subequations}
\begin{gather}
    \zeta = \frac{3}{4} \kappa \bar{r}^2,  \qquad a'=3, \qquad q' = \frac{\bar{\mu}^2 - \bar{\omega}^2}{\kappa}, \\
    \alpha' = - \sqrt{\frac{2}{3}} \frac{\bar{\omega}}{\kappa}, \qquad \beta' = \sqrt{\frac{2}{3}} \frac{\bar{\omega}}{\kappa}, \\
    \gamma' = \ell + \frac{3}{2}, \qquad \delta' = 0, \qquad \epsilon' = - \ell - \frac{1}{2},
\end{gather}
\end{subequations}
and the recursion relation coefficients read
\begin{subequations}
\begin{align}
    A_k' & = 3k (2\ell + 2k + 1) = - 3 A_k, \\
    A_{k-1}' & = \frac{2 (\bar{\omega}^2 - \bar{\mu}^2)}{\kappa} = - A_{k-1}, \\
    A_{k-2}' & = - \frac{4 \bar{\omega}^2}{3 \kappa }.
\end{align}
\end{subequations}
Again, since $A_{k-1}'$ and $A_{k-2}'$ are of the same order in $\kappa$, we drop $A_{k-2}'$. Since $\frac{A_{k-1}'}{A_{k}'} = \frac{1}{3} \frac{A_{k-1}}{A_k}$, the Taylor expansion coefficients are related by $c_k' = \frac{1}{3^k} c_k$. However, $\zeta = 3x$, and so $c_k' \zeta^k = c_k x^k$ and we recover the same Taylor expansion and the same relation to the spherical Bessel function in the $\kappa \to 0$ limit. 


\section{The P\"oschl-Teller potential}
\label{app:B}

The differential equation that governs static modes near the Buchdahl limit can be rewritten in terms of the generalized P\"oschl-Teller Hamiltonian. To see this, re-write the differential equation from Eq. \eqref{eq:2F1eq} as
\begin{align} \label{eq: hDEsimple}
    g'' + Pg' + Q = 0,
\end{align}
where $P = \frac{C}{x} + \frac{1+A+B-C}{x-1}$ and $Q = \frac{AB}{x(x-1)}$. Note that $1+A+B-C$ simplifies to $\ell + \frac{3}{2}$. Transform the field through $g(x) = \exp\left(- \frac{1}{2} \int^x P (x') dx' \right) \phi(x)$, to obtain a Schr\"odinger equation for $\phi(x)$. Change variables to $z$, given by $x = - \sinh^2 z$, and set $\phi(x) = \sqrt{dx/dz} \, \psi(z)$. The equation for $\psi (z)$ reads
\begin{equation}
    \frac{d^2 \psi}{dz^2} + \biggl( \mathcal{E} - \frac{g_s}{\sinh^2 z} - \frac{g_c}{\cosh^2 z} \biggr) \psi = 0.
\end{equation}
where $\mathcal{E} = \frac{9}{2} \bar{\mu}^2 + 48 \xi - 9$ is the energy, and $g_s$ and $g_c$ are given by $g_s = \ell ( \ell + 1 ) + 2 - 12 \xi$ and $g_c = - \ell ( \ell +  1)$. This is precisely a Schr\"odinger equation with a P\"oschl-Teller potential. However, $z$ has a finite domain, so $\psi$ has different boundary conditions compared to the standard quantum problem. Therefore, for example, quantization of bound-state  energy levels, which originates from asymptotic conditions in $|z| \rightarrow \infty$ and smoothness conditions of the wave function, do not translate directly into our system of interest. 


\section{Dynamic modes near Buchdahl}
\label{app:C}

Solving Eq.~\eqref{eq:dHyper} and taking $a \rightarrow 0$ shows that the coefficients of the spherical Bessel functions in Eq.~\eqref{eq:DynamicBuchdahl} diverge in this limit. The solution to Eq.~\eqref{eq:dHyper} with $a$ small but finite is
\begin{align} \label{eq: divergingHyper}
    h(x) & = (x-a)^{\bar{\alpha}}  \biggl[ \mathcal{A} |a|^{\bar{\lambda}} x^{\frac{1}{2}(1-\bar{\lambda})} \, {}_2 F_1 \left( \bar{\alpha}, \bar{\alpha} ; 1 - \bar{\lambda}; z \right) \notag \\ & \quad + \mathcal{B} x^{\frac{1}{2}(1 + \bar{\lambda})} \, {}_2 F_1 \left( \bar{\alpha}, \bar{\alpha} ; 1 + \bar{\lambda} ; z \right) \biggr],
\end{align}
for some constants $\mathcal{A}$, $\mathcal{B}$, and
\begin{align} \label{eq:barAlpha}
    \bar{\lambda} & = \ell + \frac{1}{2}, &%
    \bar{\alpha} & =  \frac{\bar{\omega}}{\sqrt{\kappa |a|}}, &%
    z &= - \frac{\kappa \bar{\alpha}^2 x}{\bar{\omega}^2}.
\end{align}
Next, we use the hypergeometric inversion formula
\begin{align} \label{eq: 2F1inverseFormula}
    {}_2 F_1 (A,B;C;z) &= \frac{\mathcal{A}_1}{(-z)^A} \, {}_2 F_1 \bigl( A,B_1;C_1; z^{-1} \bigr) \notag \\
    &\quad + \frac{\mathcal{A}_2}{(-z)^B} \, {}_2 F_1 \bigl( B,B_2;C_2; z^{-1} \bigr),
\end{align}
where
\begin{align}
    \mathcal{A}_1 &= \frac{\Gamma(C) \Gamma(B-A)}{\Gamma(B) \Gamma(C-A)}, &%
    \mathcal{A}_2 &= \frac{\Gamma(C) \Gamma (A - B)}{\Gamma(A) \Gamma (C - B)},
\end{align}
and
\begin{subequations}
\begin{align}
B_1 &= A-C+1, &%
C_1 &= A-B+1, \\
B_2 &= B-C+1, &%
C_2 &= B-A+1.
\end{align}
\end{subequations}
This turns Eq.~\eqref{eq: divergingHyper} into
\begin{align} \label{eq: newHyper}
    h(x) &= \mathcal{C}_1 x^{\frac{1}{2}(1 - \bar{\lambda})} \, {}_2F_1 \left( \bar{\alpha}, \bar{\alpha} ; 1 + \bar{\lambda}; - \frac{\bar{\omega}^2}{\kappa x \bar{\alpha}^2} \right) \notag \\ 
    &\quad + \mathcal{C}_2 x^{\frac{1}{2} (1 + \bar{\lambda})} \, {}_2F_1 \left( \bar{\alpha}, \bar{\alpha}; 1 - \bar{\lambda}; -\frac{\bar{\omega}^2}{\kappa x \bar{\alpha}^2} \right),
\end{align}
where
\begin{subequations}
\begin{align}
    \mathcal{C}_1 & = \frac{|a|^{-\bar{\alpha}} \Gamma(-\bar{\lambda})}{\Gamma(\bar{\alpha}) \Gamma(- \bar{\alpha})}\left[  \mathcal{B} \ \Gamma(1 + \bar{\lambda})  +  \mathcal{A} \ \Gamma(1-\bar{\lambda}) \right], \label{eq:TheC1}\\
    \mathcal{C}_2 & = \frac{\Gamma(\bar{\lambda})}{\Gamma(- \bar{\lambda})} \mathcal{C}_1.
\end{align}    
\end{subequations}
These expressions are derived using the approximation $\bar{\alpha} \pm \bar{\lambda} \approx \bar{\alpha}$. Otherwise, they would only be slightly more complicated. In the $a \to 0$ limit, $\bar{\alpha} \rightarrow \infty$, and using the property
\begin{align}\label{eq:confluentHyper}
    \lim_{a, b \to \infty} {}_2F_1 \left( a,b,c, \frac{z}{ab} \right) = {}_0F_1 \left( ; c; z \right),
\end{align}
Eq.~\eqref{eq: newHyper} is expressed in terms of ${}_0 F_1$. The latter is related to the Bessel function of first kind via
\begin{align}
    J_{\nu} (z) = \frac{(z/2)^{\nu}}{\Gamma(\nu +1)} \, {}_0F_1 \left(; \nu +1; - \frac{z^2}{4}  \right),  
\end{align}
and to the second kind via $J_{-\ell - \frac{1}{2}} (z) = (-1)^{\ell +1} Y_{\ell + \frac{1}{2}} (z)$, which is valid for integer $\ell$. Finally, $h(x)$ is written in terms of Bessel functions as
\begin{align} \label{eq: hyperBessel}
     \hspace{-0.19cm} h(x) \! = \! \mathcal{A}_{\ell} \sqrt{x} J_{\ell + \frac{1}{2}} \left( \tfrac{2 \bar{\omega}}{\sqrt{\kappa_B x}} \right) + \mathcal{B}_{\ell} \sqrt{x} Y_{\ell + \frac{1}{2}} \left( \tfrac{2 \bar{\omega}}{\sqrt{\kappa_B x}} \right)\!,
\end{align}
where
\begin{subequations}
\begin{align}
    \mathcal{A}_{\ell} & = \left( \frac{\sqrt{\kappa}}{\bar{\omega}} \right)^{\bar{\lambda}} \mathcal{C}_1 \Gamma (1 + \bar{\lambda}),  \\
    \mathcal{B}_{\ell} & =  (-1)^{\ell + 1} \left( \frac{\sqrt{\kappa}}{\bar{\omega}} \right)^{-\bar{\lambda}} \mathcal{C}_2 \Gamma(1 - \bar{\lambda}).
\end{align}    
\end{subequations}
These coefficients scale as $\mathcal{A}_{\ell}, \mathcal{B}_{\ell} \sim |a|^{-\frac{\bar{\omega}}{\sqrt{\kappa_B |a|}}}$, which diverge in the $a \to 0$ limit. 

Using the standard relationship between the ordinary and spherical Bessel functions, $h(x)$ can be expressed in terms of $j_{\ell}$ and $y_{\ell}$, ultimately resulting in Eq.~\eqref{eq:DynamicBuchdahl}.

\vfill

\end{document}